\documentclass[12pt,letterpaper]{article}

\pdfoutput=1
\usepackage{amsmath,amssymb,calc}
\usepackage{graphicx}
\usepackage{float}
\usepackage{amstext}    
\usepackage{array}    
\usepackage{color}
\usepackage{chngcntr}

\usepackage{hyperref}

\counterwithin*{equation}{section}

\newcommand\be{\begin{equation}}
\newcommand\ee{\end{equation}}
\newcommand{\bea}{\begin{eqnarray}}
\newcommand{\eea}{\end{eqnarray}}

\newcommand{\Mcal}{{\mathcal M}}

\newcommand{\nn}{\nonumber}
\newcommand{\pd}{\partial}

\def\id{\protect{{1 \kern-.28em {\rm l}}}}

\def\D{\Delta}

\def\b{\beta}

\def\k{\kappa}

\def\1{^{(1)}}
\def\0{^{(0)}}
\def\2{^{(2)}}

\def\id{\protect{{1 \kern-.28em {\rm l}}}}

\newcommand{\bra}[1]{\langle #1 |}
\newcommand{\ket}[1]{| #1 \rangle}

\newcommand*{\affaddr}[1]{#1} 
\newcommand*{\affmark}[1][*]{\textsuperscript{#1}}

\def\Cincy{\small{Department of Physics, University of Cincinnati, Cincinnati, Ohio 45221, USA}}
\def\Weizmann{\small{Department of Particle Physics and Astrophysics, Weizmann Institute of Science, Rehovot 761001, Israel}}

\begin{document}

\date{}
\title{\Large\bfseries Expansion in Higher Harmonics of Boson Stars using a Generalized Ruffini-Bonazzola Approach,\\ Part 1: Bound States}

%

\author{%
Joshua Eby,\affmark[1,]\!
	\footnote{Electronic Address: \it{joshaeby@gmail.com}} \,
Peter Suranyi,\affmark[2,]\!
	\footnote{Electronic Address: \it{peter.suranyi@uc.edu}} \,
and 
L.C.R. Wijewardhana\affmark[2,]\!
	\footnote{Electronic Address: \it{rohana.wijewardhana@uc.edu}}
\vspace{.3cm} \\
\hspace{-1cm}{\it\affaddr{\affmark[1]\Weizmann}}\\
{\it\affaddr{\affmark[2]\Cincy}}\\
}

\begin{titlepage}
\centering
\maketitle
\thispagestyle{empty}


\begin{abstract}
\noindent
 The method pioneered by Ruffini and Bonazzola (RB) to describe boson stars involves an expansion of the boson field which is linear in creation and annihilation operators. In the nonrelativistic limit, the equation of motion of RB is equivalent to the nonlinear Schr\"odinger equation. Further, the RB expansion constitutes an exact solution to a non-interacting field theory, and has been used as a reasonable ansatz for an interacting one. In this work, we show how one can go beyond the RB ansatz towards an exact solution of the interacting operator Klein-Gordon equation, which can be solved iteratively to ever higher precision. Our Generalized Ruffini-Bonazzola approach takes into account contributions from nontrivial harmonic dependence of the wavefunction, using a sum of terms with energy $k\,E_0$, where $k\geq1$ and $E_0$ is the chemical potential of a single bound axion. The method critically depends on an expansion in a parameter $\Delta \equiv \sqrt{1 - E_0{}^2/m^2} < 1$, where $m$ is the mass of the boson. In the case of the axion potential, we calculate corrections which are relevant for axion stars in the transition or dense branches of solutions. We find with high precision the local minimum of the mass, $M_{min}\approx 463\,f^2/m$, at $\Delta\approx0.27$, where $f$ is the axion decay constant. This point marks the crossover from the transition branch to the dense branch of solutions, and a corresponding crossover from structural instability to stability.
\end{abstract}

\end{titlepage}

\section{Introduction} \label{Intro}
Self-adjoint scalar particles, including axions \cite{PQ1,PQ2,Weinberg,Wilczek,DFS,Zhitnitsky,Kim,Shifman}, can form condensates known as boson stars which possibly contribute to dark matter.\footnote{For generic boson stars, see \cite{Kaup,RB,BreitGuptaZaks,CSW,SS,Friedberg,SS2,Liddel,Lee,ChavanisMR,ChavanisMR2,Kling1,Kling2}; for the specific case of axion stars, see \cite{SikivieYang,Witten,KolbTkachev,Iwazaki,BarrancoNS,TkachevFRB,BB,ESVW,Guth,Braaten,ELSW,ELSW2,WilczekASt,ChavanisPT}} There are several approaches one can use to describe these condensates. Ruffini and Bonazzola (hereafter, RB) \cite{RB} developed a relativistic approach, which solved the problem of non-interacting bosons in their own gravitational field. In the ground state, restricted to rotation-symmetric objects, a solution for the boson field was found in the form
\be \label{RB}
\Phi_{RB}=R(r)\left[\,e^{-i\,E_0\,t}\,a_0+e^{i\,E_0\,t}\,a_0^\dagger\right],
\ee
where $m$ is the mass of the axion, $E_0<m$ is its chemical potential, and $a_0$ ($a_0^{\dagger}$) is the corresponding annihilation (creation) operator. For noninteracting bosons, the number operator $\hat{N} = a_0^\dagger\,a_0$ commutes with the mass operator, 
\begin{equation} \label{MassT00}
\hat{M}=\int d^3r \,\sqrt{-g}\,T_0^0,
\end{equation}
where $T_\mu^\nu$ is the stress-energy tensor. This implies that the number of bosons is conserved.

Boson stars can only contribute to dark matter if they are sufficiently long-lived; it is thus important to distinguish the ways in which a boson star might be unstable. First, there can be solutions of the equations of motion which correspond to maxima (rather than minima) of the energy. These are unstable to either collapse or expansion, depending on how they are perturbed; we refer to this as \emph{structural instability}.\footnote{In the context of astrophysical bodies, this is often referred to as \emph{gravitational stability}; however, there are situations in which gravity can be decoupled and yet a similar instability arises through other interactions. For this reason, we have chosen to use a more generic term.} Second, if the constituent bosons are real scalars, their number is not conserved, and such boson stars can decay through self-interactions \cite{ESW,Braaten2016,MTY,EMSW} (or potentially to Standard Model particles, if the two sectors are coupled); if the decay rate is sufficiently fast, we call this \emph{decay instability}.

The RB operator of eq. (\ref{RB}) is a solution to the operator Klein-Gordon (KG) equation for non-interacting bosons. The same operator was applied to interacting boson stars with the axion potential first by Barranco and Bernal~\cite{BB}. Shortly after, we undertook a systematic study of dilute axion stars~\cite{ESVW}, which are defined as having low central density as characterized by weak field strength, $X(r)=2\,\sqrt{N}\,R(r)\,/\,f  \ll 1$, where $f$ is the decay constant of axions and $N$ is the number of axions. We found that all physical quantities depended on an important scaling parameter 
\begin{equation}
 \Delta =\sqrt{1-\frac{E_0{}^2}{m^2}},
\end{equation}
and we calculated the most important properties (mass, radius, density) of axion stars under the assumption of small binding energy. 

Our final aim is a systematic study of boson stars, including those with relatively large binding energy. One reason for this is the application to axions, to accurately describe all possible axion star states, which can exist in a number of distinct forms. In the {\em dilute} axion star region, the scaling parameter $\Delta$ satisfies $\Delta \lesssim \mathcal{O}(f\,/\,M_P)$, where $M_P$ is the Planck mass~\cite{ESVW}. When $\Delta$ is increased above this point, the metastable dilute axion star becomes unstable as the mass of the star reaches a maximum value.  However, a {\em dense state} \cite{Braaten} also exists in the range $\Delta = \mathcal{O}(1)$, with much smaller radius $R_a\sim m^{-1}$. The well-known methods, most prominently the nonrelativistic Gross-Pitaevskii (GP) and RB methods, break down in this region because the field values and the binding energy are both large, giving rise to important relativistic corrections.

The central problem is that as $\Delta$ increases, the ansatz (\ref{RB}) becomes more and more unreliable. The leading order KG equation for the wavefunction $R(r)$, which is equivalent to the GP equation in the nonrelativistic limit, becomes inadequate. Unlike the model discussed in~\cite{RB}, in the presence of self-interaction terms the operator equation of motion is not solved exactly by (\ref{RB}); self-interactions give rise to nontrivial contributions from special relativity at large binding energies. In particular, it becomes inappropriate to assume the field is dominated by a single harmonic time dependent expression. Rather, one must include in the wavefunction terms corresponding to higher-order harmonics, $e^{i\,k\,E_0\,t}$ with $k>1$; this point was made recently in \cite{WilczekASt}, and a first attempt to do so was presented in the context of the GP equation there. 

In this note, we point out that the RB method must be amended by relativistic corrections in interacting theories. With these considerations in mind, we present our work based on the RB formalism, in which we use a consistent expansion in $\Delta$ that takes these higher harmonics into account; we refer to this method as a Generalized Ruffini-Bonazzola expansion, or GRB. Such an expansion is possible in part because off-diagonal elements in the operator KG equation are proportional to increasingly large powers of $\Delta<1$. In this framework we can accurately describe scattering states as well, allowing a consistent description of boson star decay, which we leave for Part 2 of this work.

This paper is organized as follows. After describing the relevant equations in the most general terms in Section \ref{BEC}, we present the GRB approach for calculating bound state wavefunctions in Section \ref{SolBS}; we do so in a model-independent way by evaluating expressions in terms of a general self-interaction potential $V(\Phi)$, and we also use a $\Phi^4$ interaction as an illustrative example. We specialize to the case of the axion potential in Section \ref{Axions}, where we analyze the leading relativistic corrections in GRB on the structure of axion stars. We test the structural stability of all of our solutions using a perturbation analysis in Section \ref{Axions} as well. We conclude in Section \ref{Conclusion}.

We will use natural units throughout, where $\hbar = c = 1$.

\section{Bosonic Condensates} \label{BEC}

In general, a boson star is supported by a balance of kinetic, gravitational, and self-interaction forces. In this work, unless otherwise specified, we will neglect the effect of the self-gravity of the boson star, though its inclusion should not in principle affect the formalism in any important way. Introducing weak gravity would not be difficult, but the notations are more cumbersome. On the other hand, in some applications no stable state exists without the inclusion of gravity, e.g. for repulsively interacting bosons. To make the presentation clearer, for now, we will ignore this complication.

We begin with the action of a real scalar field, which is
\be \label{action}
 S=\frac{1}{2}\int d^3r\,dt \left[\pd_\mu\Phi\, \pd^\mu \Phi -2\,V(\Phi)\right].
\ee 
The corresponding Hamilton operator (integral of $T_0^0$) is
 \be\label{H}
H[\Phi]=\frac{1}{2}\int d^3r\, \left[\Phi'{}^2+\dot\Phi^2 +2\, V(\Phi)\right].
\ee 
We will begin by working with a general potential $V(\Phi)$ to make some of the results more transparent, as well as more general. The axion instanton  potential 
\begin{equation} \label{Vaxion}
 V_a(\Phi)=m^2\,f^2 \left[1-\cos\left(\frac{\Phi}{f}\right)\right],
\end{equation}
is just one particular choice.
  
The  quantum variational equation of motion for $\Phi$ derived from eq. (\ref{action}) is
\be \label{KGPhi}
KG[\Phi]=-\ddot\Phi+\nabla^2\Phi-V'(\Phi)=0.
\ee
The full spectrum of states denoted by $\Phi$ can be written as
\begin{equation}
 \Phi = \Phi_b + \Psi,
\end{equation}
where $\Phi_b$ ($\Psi$) denotes the bound (scattering) state contribution.
In $\Phi_b$, we will assume perfect $N$-particle condensate of particles in the ground state, which is valid at zero temperature. Then we can expand the operator KG equation using a collection of these $N$-particle states, denoted
\begin{equation}
 \ket{N} = \frac{a_0^{\dagger N}}{\sqrt{N!}}\ket{0}.
\end{equation}
 
The equation of motion for the bound state operator $\Phi_b$ is the operator KG equation, $KG[\Phi_b]=0$.  $\Phi_b$ is a regular, self-adjoint function of the absorption and emission operators $a_0(t)= a_0\, e^{-i\,E_0\,t}$ and $a^\dagger_0(t)=a_0^\dagger\, e^{i\,E_0\,t}$ of bound state bosons, which have eigenenergy $E_0<m$. When $V(\Phi) = m^2\,\Phi^2/2$ (no self-interactions), the expansion of eq. (\ref{RB}) is exactly a solution of the full operator equation (\ref{KGPhi}) for an appropriately chosen wavefunction $R(r)$. However, a nontrivial self-interaction potential spoils this conclusion, as we will explain below. In the next section, we describe our Generalized Ruffini-Bonazzola (GRB) method and show that it provides an iterative solution to the interacting field theory for boson stars.

The scattering state operator $\Psi$ produces quanta with energy $E_p>m$ and which are thus not bound to the boson star. These quanta give rise to a decay rate for boson stars formed from Hermitian scalars, as is the case for axions. In \cite{ESW,EMSW}, we parameterized these scattering states by free spherical waves; this is altogether appropriate far from the star and is only perturbed by binding energy contributions inside the star, which are small when the star is dilute. This simplification breaks down for more strongly bound boson stars. We will show how scattering states can be included self-consistently within our GRB analysis in Part 2 of this work.

\section{Solution for the Bound States} \label{SolBS}

\subsection{General Framework}
In this section we analyze the bound state contributions to the KG equation, eq. (\ref{KGPhi}). Satisfying the operator KG equation for the bound state is equivalent to satisfying the simultaneous equations
\begin{equation} \label{Mk}
 \Mcal_k = \bra{N}KG[\Phi_b]\ket{N-k} = 0
\end{equation}
for all integers $k$. We choose to focus on $k>0$, for a number of reasons. First, expanding the field around $\Phi_b=0$ justifies using $\Mcal_0=0$. Further, the equations for $-N<k<0$ are equivalent to those of $0<k<N$, so we can safely ignore $k<0$. We will see that the parameterization of eq. (\ref{Mk}) gives $N\ggg1$ correction terms to the nonrelativistic limit, an approximation which will be applicable far beyond any practical application.

In the case of a trivial interaction potential $V(\Phi)=m^2\,\Phi^2\,/\,2$, the operator equation
\be \label{noninteracting}
-\pd_t{}^2\Phi-m^2\,\Phi+\nabla^2\Phi=0,
\ee
is solved by the spherically symmetric, self-adjoint, RB solution in eq. (\ref{RB}). In this case the only nontrivial equation in the collection (\ref{Mk}) is $\Mcal_1=0$, which can be solved for the single wavefunction $R(r)$; all higher-order terms $\Mcal_{k>1}=0$ are satisfied trivially.

However, when we replace the mass term by a general (non-purely-quadratic) potential $V(\Phi)$, then the solution becomes more complicated.  Substituting (\ref{RB}) into  (\ref{KGPhi}) cannot solve the operator equation, because upon substitution into the potential term, we generate uncompensated terms with higher powers of the creation and annihilation operators, $a_0^\dagger$ and $a_0$. One can still solve the leading-order equation $\Mcal_1=0$, but we emphasize that using (\ref{RB}) in a theory with self-interactions \emph{cannot solve the operator equation of motion}. Each equation $\Mcal_k=0$ is a different function of the field $\Phi$ which cannot be simultaneously solved by a single wavefunction $R(r)$. Using (\ref{RB}) becomes even more questionable if we calculate matrix elements relevant to boson star decay, such as $\bra{N}\Phi_b^3\,\Psi\ket{N-3,p}$, where $\ket{N-3,p}$ denotes the direct product of a bound state of $N-3$ bosons and a single scattering state axion of momentum magnitude $p$ \cite{ESW,EMSW}.

The simplest solution is to generalize $\Phi_{\rm RB}$, such that it contains additional terms with higher powers of of $a_0$ and $a_0^\dagger$. Thus we introduce the Generalized RB (GRB) operator:
\begin{align}\label{GRB}
\Phi_b  &=\sum_{k=1}^\infty(\Phi_{k}^-+\Phi_{k}^+) \nn \\
      &\equiv\sum_{k=1}^\infty\,R_{k}(r)\,\left[e^{-i\,k\,E_0\,t}\,a_0^{k}
	    +e^{i\,k\,E_0\,t}\,a_0^{\dagger\,k}\right].
\end{align}
It is easy to see that the $k=1$ contribution is the original RB ansatz, but the full GRB expansion includes an infinite series of higher harmonics. There is no need to introduce terms with mixed products of $a_0$ and $a_0^\dagger$, because in expectation values in a condensate at large $N$, the commutator $[a_0^\dagger,a_0]$ can be neglected and $\hat N=a_0^\dagger a_0$ can be replaced by its eigenvalue $N$. In the large $N$ limit, terms containing mixed products can be reduced to terms with powers of $a_0$ or powers of $a_0^\dagger$ only.

The set of equations (\ref{Mk}), along with the GRB solution (\ref{GRB}), contain all relevant corrections coming from special relativity. By comparison with the GP equation, we can see that these corrections come in two forms: first, the expansion of (\ref{GRB}) takes into account the interference of higher-order harmonics in the field, usually neglected in the nonrelativistic limit; and second, the KG equation contains a second derivative in time. Note that this second time derivative term contains important relativistic corrections which do not appear the GP equation, and cannot be neglected beyond the nonrelativistic limit.

In this work, we will represent time derivatives explicitly in terms of the chemical potential $E_0$, as in the original works on boson stars \cite{Kaup,RB}. Alternatively, one can exchange higher-order time derivatives for spatial derivatives using the equations of motion \cite{Kiefer,Giulini,Guth2,Banerjee}. Such choices merely reparameterize the problem, and in the analysis of static bound states, any physical results should be equivalent.

Using the GRB operator (\ref{GRB}), we can rewrite the KG expectation values (\ref{Mk}) as
\begin{equation} \label{Mk2}
 e^{-i\,k\,E_0\,t}\Mcal_k = \left[(k\,E_0)^2 - m^2 + \nabla^2\right]R_k
	  - e^{-i\,k\,E_0\,t} \bra{N} V_{int}'(\Phi_b)\ket{N-k} = 0,
\end{equation}
where we have defined the interacting part of the potential $V_{int}(\Phi) = V(\Phi) - m^2\,\Phi^2/2$. Now to solve (\ref{Mk2}) by finding the coefficient functions $R_{k}(r)$ may seem like an impossible task for a general potential $V_{int}(\Phi)$. To derive equations for coefficients $R_{k}$ one has to simultaneously satisfy $\Mcal_{k}=0$ for all $k\geq1$. We illustrate how this procedure can indeed lead to a tractable expansion through the use of an example.

\subsection{Example: Quartic Potential} \label{quarticSec}

Let us consider a quartic potential
\begin{equation} \label{quartic}
 V_{int}(\Phi_b) = \frac{\lambda}{4!}\Phi_b^4.
\end{equation}
It gives a cubic term in $V'(\Phi_b)$, which contributes to $\Mcal_{k+1}$ any term which is a product of $R_{m+1}R_{\ell+1}R_{n+1}$, as long as $k-1/2=\pm(\ell-1/2)\pm(m-1/2)\pm(n-1/2)$. Because the potential is an even function of $\Phi_b$ in this case, without loss of generality we can set $\Mcal_{2j}=0$; however, the odd-numbered equations $\Mcal_{2j+1}=0$ still constitute a very large set, which must be solved for many coefficient functions $R_{2j+1}$.

To see how such this procedure can be made tractable, consider the restricted expansion of eq. (\ref{GRB}) up to $k_{\rm max}=3$. For the quartic potential in eq. (\ref{quartic}), this implies we will have to solve for the coefficient functions $R_1$ and $R_3$. Substituting the GRB bound state expansion (\ref{GRB}) into (\ref{Mk2}) we obtain the following leading order equations, $\Mcal_1=0$ and $\Mcal_3=0$, for $k_{\rm max}=3$:
 \begin{align}
 -\Delta^2\,m^2\,Z_1(r) + \nabla^2\,Z_1(r) &- \frac{\lambda}{8}
	  \left[Z_1(r)^3 + Z_1(r)^2\,Z_3(r) 
		  + 2\,Z_1(r)\,Z_3(r)^2\right] = 0	\label{Ze1}\\
  (9\,E_0{}^2-m^2)\,Z_3(r) + \nabla^2\, Z_3(r) &- \frac{\lambda}{24}
	  \left[Z_1(r)^3 + 6\,Z_1(r)^2\,Z_3(r)
		  + 3\,Z_3(r)^3\right]  = 0			\label{Ze3}
  \end{align}
  where we introduced the rescaled wavefunctions 
  \begin{equation} \label{Zk}
   Z_k(r) = 2\,N^{k/2}\,R_k(r), \qquad (k\geq1).
  \end{equation}
  
Following our previous work \cite{ESVW}, we can rescale both the leading order wavefunction and coordinate with a power of $\Delta$ corresponding to their operator dimension, $x=\Delta\,m\,r$ and $Z_1(r)\propto\Delta$. Clearly a solution of eq. (\ref{Ze3}) will require $Z_3(r) = \mathcal{O}(Z_1{}^3) \propto \D^3$; when $\Delta\ll1$, this represents a small correction to the leading boson star wavefunction $Z_1$. This identification is crucial, and gives rise to a well-defined expansion in $\Delta$. To make explicit the truncation of the series in $\D$, we will use the notation that $\Mcal_{k}^{(n)}$ denotes $\Mcal_k$ expanded to $\mathcal{O}(\D^n)$. 

The leading order expression of eq. (\ref{Ze3}) is $Z_3 \approx \lambda\,Z_1{}^3/(192\,m^2)$, which is $\mathcal{O}(\Delta^3)$. At this order, $Z_3$ does not couple to $Z_1$ in eq. (\ref{Ze1}) at all. To be sensitive to a nontrivial backreaction of $Z_3$ on the leading order wavefunction $Z_1$, we must keep terms in eq. (\ref{Ze1}) at least to order $\D^5$; the result for $\Mcal_1^{(5)}=0$ is
 \begin{equation} \label{Z1toD5}
  0 = -\Delta^2\,m^2\,Z_1(r) + \nabla^2\,Z_1(r) - \frac{\lambda}{8}
	  \left[Z_1(r)^3 + \frac{\lambda}{192}Z_1(r)^5 \right].
 \end{equation}
 Observe that although the potential for $\Phi_b$ contained at most a quartic term, corrections generate perturbatively an effective interaction potential for $Z_1$
 \begin{equation}
  V_{eff} = \frac{\lambda}{2^6}\,Z_1{}^4 
	+ \frac{\lambda^2}{3^2\times2^{11}}\,Z_1{}^6,
 \end{equation}
 which includes a $Z_1{}^6$ interaction. This potential is relevant for calculating the $3\to3$ scattering cross section in a boson star, through a Feynman diagram which has an internal axion line with energy $3\,E_0$.
 
 Taking into account the backreaction of $Z_3$ on $Z_1$ can be thought of as integrating out fast oscillating modes with frequency $3\,E_0$. This is the view taken by \cite{MTY}, who integrate this mode out to find the effective potential for the nonrelativistic component of the field (proportional to what we call $Z_1$). We have checked that the effective potential generated by our method agrees with their result up to $\mathcal{O}(Z_1^6)$.
   
 We can extend this procedure to ever higher corrections $Z_k(r)$ to the wavefunction. In particular, the fact that
  \begin{equation}
   Z_k(r) \propto \Delta^k
  \end{equation}
 holds is what allows for a consistent expansion in $\D$. One can solve (\ref{KGPhi}), considering more and more equations  $\Mcal_1=0,\,\Mcal_2=0,...$  and finding wave functions $R_1,\,R_2,...$  one by one, in ever increasing precision, as $R_{k}\propto \D^k$ will rapidly decrease with $k$.
 
 It is important to emphasize the following fact: The above described procedure implies that GRB is {\em not merely an ansatz}; it provides an iterative solution to the operator equation of motion (\ref{KGPhi}). The leading order approximation is RB, but higher-order corrections are organized as a series in the parameter $\D<1$. GRB, with a cutoff at $k=k_{\rm max}$,  solves the operator equation of motion approximately to $\mathcal{O}(Z_1{}^{k_{\rm max}})\simeq \mathcal{O}(\Delta^{k_{\rm max}})$.
 
 This method can be used to systematically study strongly-bound boson stars, like the dense axion stars considered recently in the literature \cite{Braaten,ELSW,ELSW2,WilczekASt,ChavanisPT}. Such states correspond to large values of $\Delta$, even up to $\Delta=\mathcal{O}(1)$, but since $\Delta<1$ always holds, the solutions can still be found to ever higher precision using our expansion method and the GRB procedure presented here. We will set up a minimal example of this application in Section \ref{Axions}.
 
 Note also that equations very similar to (\ref{Ze1}) and (\ref{Ze3}) were derived in the nonrelativistic analysis of axions by \cite{WilczekASt}. These authors emphasize that, for dense or strongly-bound stars, the ``single harmonic'' method (i.e. the leading order RB or GP equation) breaks down, and an expansion in higher harmonics is needed. Here we have presented just such an analysis, which includes corrections at all orders in special relativity, and which can be solved self-consistently as a power series in $\Delta$. 
 
 A further strength of our method is that we can include also scattering state contributions. These states couple to the bound states and give rise to decay processes of boson stars, and in this method we can calculate the rates for specific processes at each order in $\Delta$, as we will describe in a future publication.

\section{Application to Axion Stars} \label{Axions}

The framework we outlined above is generic, and can be applied to any boson star. Nonetheless, this work is in part motivated by an attempt to characterize all possible bound states of axion particles. In this section, we specialize to the case of the axion instanton potential, eq. (\ref{Vaxion}).

The case of axion stars at very weak binding $\Delta\ll1$ is well understood \cite{ChavanisMR,ChavanisMR2,Kling1,Kling2,BB,ESVW,Guth,WilczekASt}. Those states which have $\D\lesssim f/M_P$ are metastable, whereas structurally unstable states exist for $f/M_P \lesssim \D \ll 1$; these have become known as the ``dilute'' and ``transition'' branches of axion stars, respectively. The critical point $\D_c\sim f/M_P$ is the position of the maximum mass of weakly bound axion stars, which is $M_{max} \sim M_P\,f/m$ \cite{ChavanisMR,ChavanisMR2,ESVW}. As we have shown previously \cite{ESW}, the effective gravitational coupling is ${\k\equiv 8\,\pi\,f^2 / (M_P{}^2\,\D^2)}$, and so this critical point also signifies that the gravitational interaction decouples on the transition branch, when $\k\ll1$ (see also Section \ref{ASTtoX1}). A similar point was made recently by \cite{WilczekASt}.

However, at even larger values of $\Delta=\mathcal{O}(1)$, the leading order equation of motion has another solution which is expected to be structurally stable, dubbed a ``dense axion star''; such solutions were originally found through numerical integration of the GP equation \cite{Braaten}, but are also accessible using a variational method \cite{ELSW,ELSW2}. Analysis of these dense states has since become a strong driver of recent axion star literature \cite{WilczekASt,ChavanisPT}. One concerning fact is that such states correspond to large field values, giving rise to large corrections to the GP and RB methods. Calculations of dense axion star properties using these methods rests on a somewhat unstable foundation.

It is important to point out that particle number $N$ is not a good quantum number for axions, as they are real bosons. If axions are produced in coherent states, this would also imply a nonzero spread in the particle number distribution. Nonetheless, when $\Delta\ll1$, the distribution is strongly peaked at the mean value of $N = \langle N\rangle$, and so the use of $\ket{N}$ states to describe such axion stars is appropriate. At masses smaller than the maximum $M_{max}=m\,N_{max}$, a single value of $N$ corresponds to a state on both the dilute and transition branches of solutions. In the nonrelativistic GP formulation, there is also a corresponding state on the dense branch \cite{ELSW}, though the dispersion is much more significant on this branch.

Related to the use of $N$ as a quantum number is the decay rate of axion stars through number-changing self-interactions. We have shown previously \cite{ESW,EMSW} that in limit $\Delta\ll1$, the decay rate of dilute or transition stars is strongly suppressed by a factor of roughly $e^{-1/\Delta}\lll1$. This rate can become important, for example, in the case of collapsing dilute axion stars \cite{ELSW,Tkachev2016,MarshCollapse}. The question arises, whether dense axions stars are also stable against decay. To examine that question we need to extend the investigation of~\cite{ESW,EMSW} to larger values of $\Delta$. As a first step, this work will advance a precise characterization of the bound state configurations; in a forthcoming companion work, we will show how to couple them to scattering states which give rise to decay.

\subsection{Leading Order: \texorpdfstring{$\mathcal{O}(\Delta^3)$}{}; Dilute Axion Stars}
\label{ASTtoX1}

At leading order, GRB reduces to the RB ansatz in eq. (\ref{RB}). For the axion potential, eq. (\ref{Vaxion}), this was used to analyze the axion potential by various groups previously \cite{BB,ESVW}, though for completeness we review it here. The leading order equation of motion is $\Mcal_{1}^{(3)}=0$, which is equivalent to the GP equation considered by numerous authors \cite{ChavanisMR,ChavanisMR2,Braaten,ELSW,ELSW2,WilczekASt,ChavanisMR2,Kling1,Kling2}. This application should be valid at the very smallest values of $\Delta\ll 1$. The equation is
\begin{equation}
 -\Delta^2 m^2 X_1(r) + \nabla^2 X_1(r) + \frac{m^2}{8}X_1(r)^3
      - m^2\,\frac{8\,\pi\,f^2}{M_P{}^2\,\Delta^2}\,b(r)\,X_1(r)	= 0,
\end{equation}
where $X_k(r)=Z_k(r)/f$, with $Z_k$ defined in eq. (\ref{Zk}),\footnote{This matches the notation introduced in \cite{ESVW}, and also Section \ref{SolBS}.} and $b(r)$ a rescaled Newtonian gravitational potential (defined below). It is important in this regime to include the effect of gravity, as well as the quartic self-interaction coupling $\lambda = -m^2/f^2$. Following our previous work \cite{ESVW,ESW} and the discussion in Section \ref{SolBS}, we rescale the coordinate and wavefunction by
\begin{align}\label{Yintro}
  r&=\frac{x}{m\,\Delta},\nn\\
  X_k(r)&=\Delta^k\,Y_k(x), ~~~~ (k\geq1)
\end{align}
which gives the simple expression
 \begin{equation} \label{diluteEOM}
 \nabla_x^2\,Y_1(x) - Y_1(x) + \frac{1}{8}Y_1(x)^3 
      - \kappa\,b(x) Y_1(x) = 0,
\end{equation}
where the gravitational coupling $\kappa = 8\pi\,f^2/(M_P{}^2\,\D^2)$, and the gravitational potential is
\begin{equation}
 b(x) = -\frac{1}{8\pi}\int d^3x'\frac{Y_1(x')^2}{|\vec{x}-\vec{x}'|}.
\end{equation}

\begin{figure}[t]
 \centering
 \includegraphics[scale=.64]{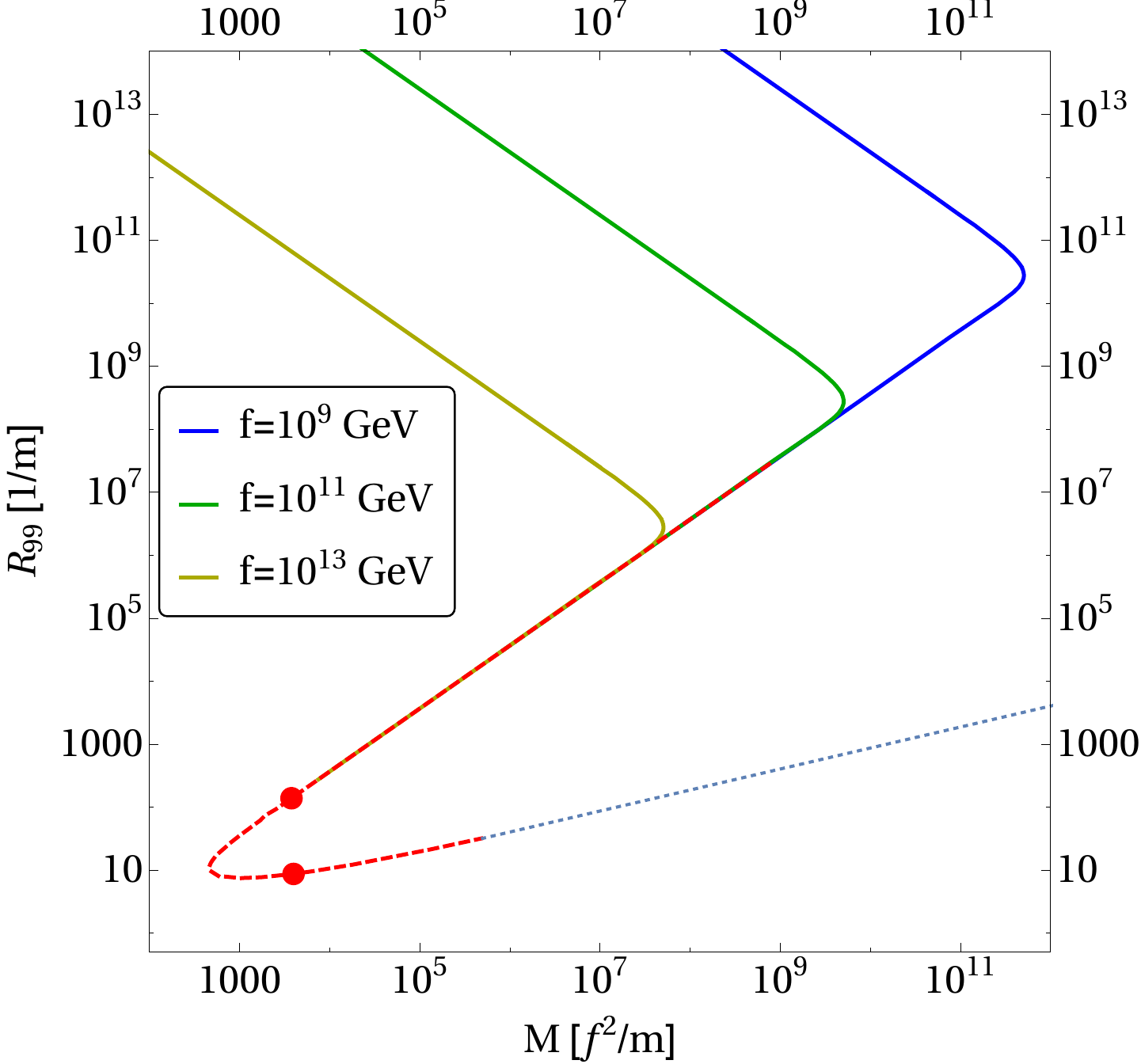}
 \caption{The mass, calculated using eq. (\ref{MassTot}), and radius $R_{99}$, of axion stars. The solid lines represent solutions of eq. (\ref{diluteEOM}), the leading order GRB equation for axion stars in the weak binding limit $\Delta\ll1$, for decay constant $f=10^9,10^{11},10^{13}$ GeV (blue, green, and yellow respectively). The dashed red line connecting the transition region and the dense branch is calculated using the GRB expansion up to $\mathcal{O}(\Delta^5)$, eqs. (\ref{Y1}), (\ref{Y3}), and (\ref{Y5}); the endpoint of this curve represents $\Delta=0.69$. Points along the blue, dotted line \emph{are not solutions}; this line instead represents the assumption that $M\sim R^3$ at even larger values of $\Delta$ on the dense branch. The upper circle, on the transition branch, corresponds to the left panel of Figure \ref{DenseSolutions} and the value $\Delta=0.02$; the lower circle corresponds to the right panel and the value $\Delta=0.6$.}
 \label{Mass_Rad}
\end{figure}

The solutions of eq. (\ref{diluteEOM}) are well known \cite{ChavanisMR,ChavanisMR2,ESVW,WilczekASt}; the mass is calculated by eq. (\ref{MassT00}) taken to the relevant order of $\D$, and following the conventional notation, we define the size of an axion star as $R_{99}$, the radius inside which $0.99$ of the mass is contained. We depict these results by the solid blue, green, and yellow lines in Figure \ref{Mass_Rad} for different choices of axion decay constant $f$. The maximum mass in this region is given approximately by $M_{max} \approx 10\,M_P\,f/m$, which also marks the transition to $\Delta \gtrsim f/M_P$ (at smaller radii), at which the gravitational coupling $\kappa\ll1$ implies that gravity decouples. These results have been reported numerous times.

\subsection{Corrections from \texorpdfstring{$X_3$}{} at \texorpdfstring{$\mathcal{O}(\Delta^3)$}{}}
\label{ASTtoX3}

We now compute the leading-order corrections in the GRB equations, which arise at $\mathcal{O}(\Delta^3)$; solutions at this order depend on two wavefunctions $R_1$ and $R_3$ in eq. (\ref{GRB}), and we can truncate the potential at $\mathcal{O}(\Phi^4)$.\footnote{As we described in Section \ref{quarticSec}, the axion potential carries only even powers of $\Phi$, the expansion of eq. (\ref{GRB}) includes only odd powers of $a_0$ and $a_0^\dagger$.} At large enough values of $\Delta$ (where corrections at this order can be important), gravity can be safely ignored. Neglecting the effect of gravity and truncating the potential at $\mathcal{O}(\Phi^4)$ corresponds to solutions in the transition region. Although these states are known to be unstable to collapse under perturbations \cite{ELSW,ELSW2,Khlopov}, they nonetheless serve as a simple illustration of the GRB method.

Because we truncate at $\mathcal{O}(\Phi^4)$, we may simply use the results of Section \ref{SolBS} to find the equations of motion describing the bound state; the two equations needed for finding $R_1$ and $R_3$ are the two ``least off-diagonal'' matrix elements of (\ref{KGPhi}). The result for $\Mcal_{1}^{(3)}=0$ and $\Mcal_{3}^{(3)}=0$ is
 \begin{align}\label{X1X3}
  -\Delta^2 m^2 X_1 + \nabla^2 X_1 + \frac{m^2}{8}X_1^3
		&= 0, \nn \\
  8\,X_3 + \frac{1}{24}X_1^3 &= 0.
 \end{align}
Note that these equations were also derived using a nonrelativistic analysis in \cite{WilczekASt}, although our method offers the quantitative advantage of a systematic expansion in a small parameter. 
 
 Next we can rescale the fields and the coordinates, as usual \cite{ESVW,ESW,EMSW} using eq. (\ref{Yintro}), and solve (\ref{X1X3}) by iteration, to get in leading order of $\Delta$
\begin{align}
 -Y_1+\nabla_x{}^2\,Y_1+ \frac{1}{8}Y_1^3=0,\label{Y1final}\\
 8\,Y_3+\frac{1}{24}Y_1{}^3=0.\label{Y3final}
 \end{align}
Eq. (\ref{Y1final}) has a {\em unique solution} for $Y_1(x)$, such that it is regular at and near the real axis, even, and satisfies $Y_1(x)<c_0\, \Delta \, e^{-x}$ \cite{ESVW}. The boundary condition giving rise to the correct behavior is $Y_1(0)=12.2679$, and at large $x$ we have $Y_1(x)\sim e^{-x}\,/\sqrt{x}$. Using (\ref{Y3final}), $Y_3(x)$ is also determined exactly at this order in $\D$. However, at this order there is no backreaction of $Y_3$ on $Y_1$, and correspondingly, no qualitatively different results compared to Section \ref{ASTtoX1}.

\subsection{Corrections at \texorpdfstring{$\mathcal{O}(\Delta^5)$}{}; Dense Axion Stars}
\label{ASTtoX5}

Dense axion stars \cite{Braaten} have large binding energies \cite{ELSW,WilczekASt}, which impiles $\Delta = \mathcal{O}(1)$ can be attained. This is precisely when corrections to the leading order equations become very important. Of course in principle, as $\Delta\to1$ the GRB expansion becomes less and less effective, as it is an expansion in $\Delta$. 

Nonetheless, we can approximate the full solutions by taking the minimum number of terms in the potential necessary to describe the dense configuration. As argued in \cite{ELSW,ELSW2}, one needs to include terms in the potential at least to $\mathcal{O}(\Phi^6)$, which is the first repulsive self-interaction term. In that case the energy is bounded from below. For the GRB parameterization, this means including wavefunctions up to $R_5$, and corrections up to $\mathcal{O}(\Delta^5)$.

We describe here how to approximate the dense axion star in this framework. Now the relevant equations of motion for the bound state come from eq. (\ref{Mk}) with $k=1,3,5$:
\begin{align}
 -m^2\,\Delta^2\,X_1 + \nabla^2 X_1 + \frac{m^2}{8}\left(X_1{}^3 
	    + X_1{}^2\,X_3 - \frac{1}{24}X_1{}^5\right) &= 0 
		  \label{X1} \\
 (9E_0{}^2-m^2)\,X_3 + \nabla^2 X_3 + \frac{m^2}{24}\left(X_1{}^3 
	    + 6X_1{}^2\,X_3 - \frac{1}{16}X_1{}^5\right) &= 0  
		  \label{X3} \\
 (25E_0{}^2-m^2)\,X_5 + \nabla^2 X_5 
	    + \frac{m^2}{8}\left(X_1{}^2\,X_3 
	    - \frac{1}{240}X_1{}^5\right) &= 0, \label{X5}
\end{align}
In the spirit of an iterative solution, we can simplify the equations by replacing $X_3$ everywhere, except in the first term of (\ref{X3}), by the leading-order expression in (\ref{X3}): $X_3 = -X_1{}^3 / 192 + \mathcal{O}(\Delta^5)$. At this order, $X_5 = \mathcal{O}(\D^5)$ does not backreact on $X_1$ and does not contribute to the mass. In terms of the rescaled wavefunctions $X_k = \Delta^k\,Y_k$ and coordinate $x = r/m\,\Delta$, we calculated up to $\mathcal{O}(\Delta^5)$ the result for $\Mcal_{1}^{(5)} = 0$, $\Mcal_{3}^{(5)} = 0$, and $\Mcal_{5}^{(5)} = 0$, which is
\begin{align}
 \nabla_x^2\,Y_1 &-Y_1 + \frac{1}{8}\left(Y_1{}^3 
	    - \frac{3\,\Delta^2\,Y_1{}^5}{64}\right) = 0 \label{Y1} \\
 Y_3 &= -\frac{1}{8}\left[\left(1+\frac{9\,\Delta^2}{8}\right)\frac{Y_1{}^3}{24}
	    - \frac{\nabla_x^2\left(Y_1{}^3\right)}{192} - \frac{\Delta^2\,Y_1{}^5}{256}
		  \right] \label{Y3} \\
 Y_5 &= -\frac{Y_1{}^5}{20480}. \label{Y5}
\end{align}

\begin{figure}[t]
 \centering
 \includegraphics[scale=.54]{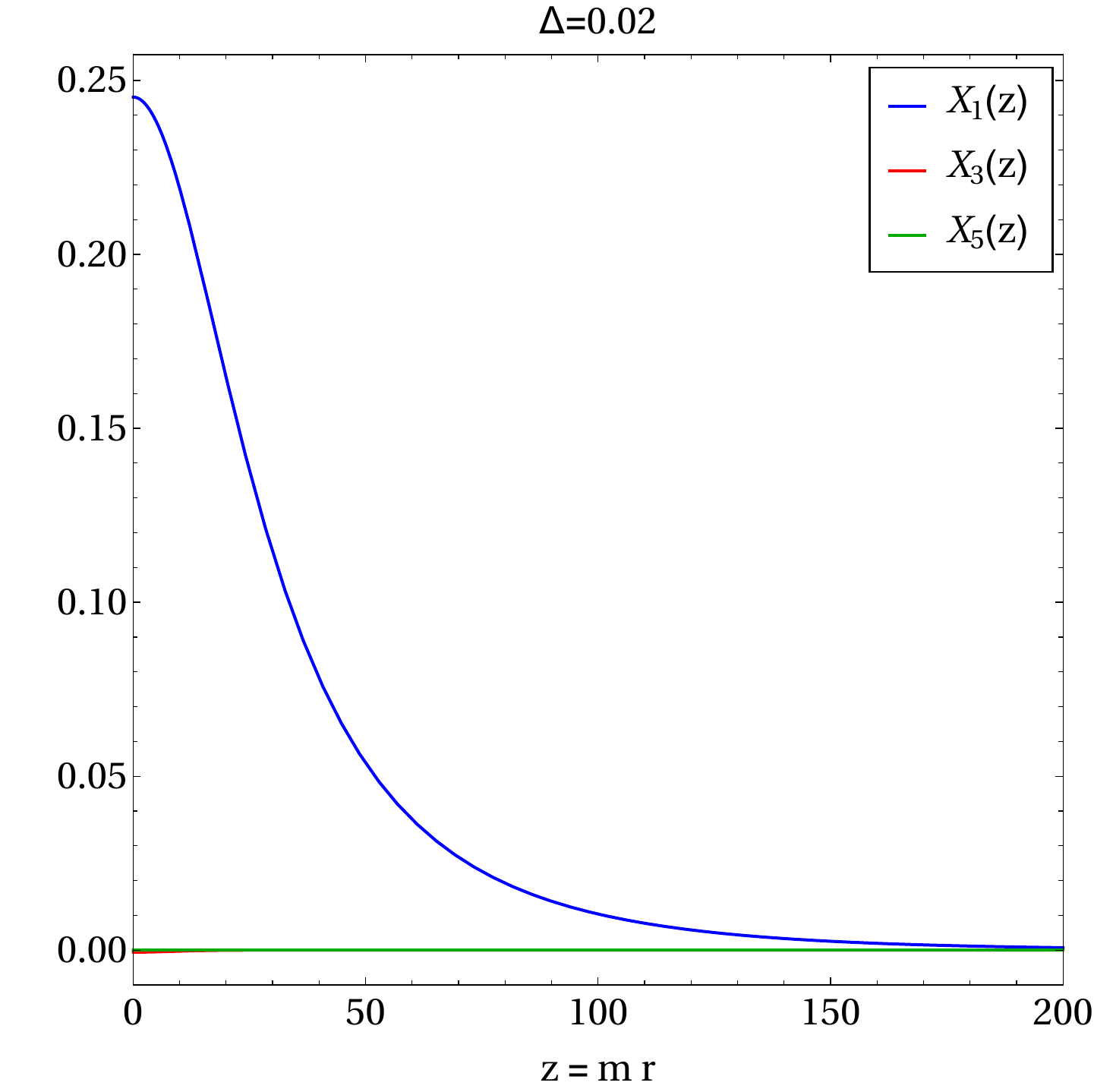} ~~~
 \includegraphics[scale=.5]{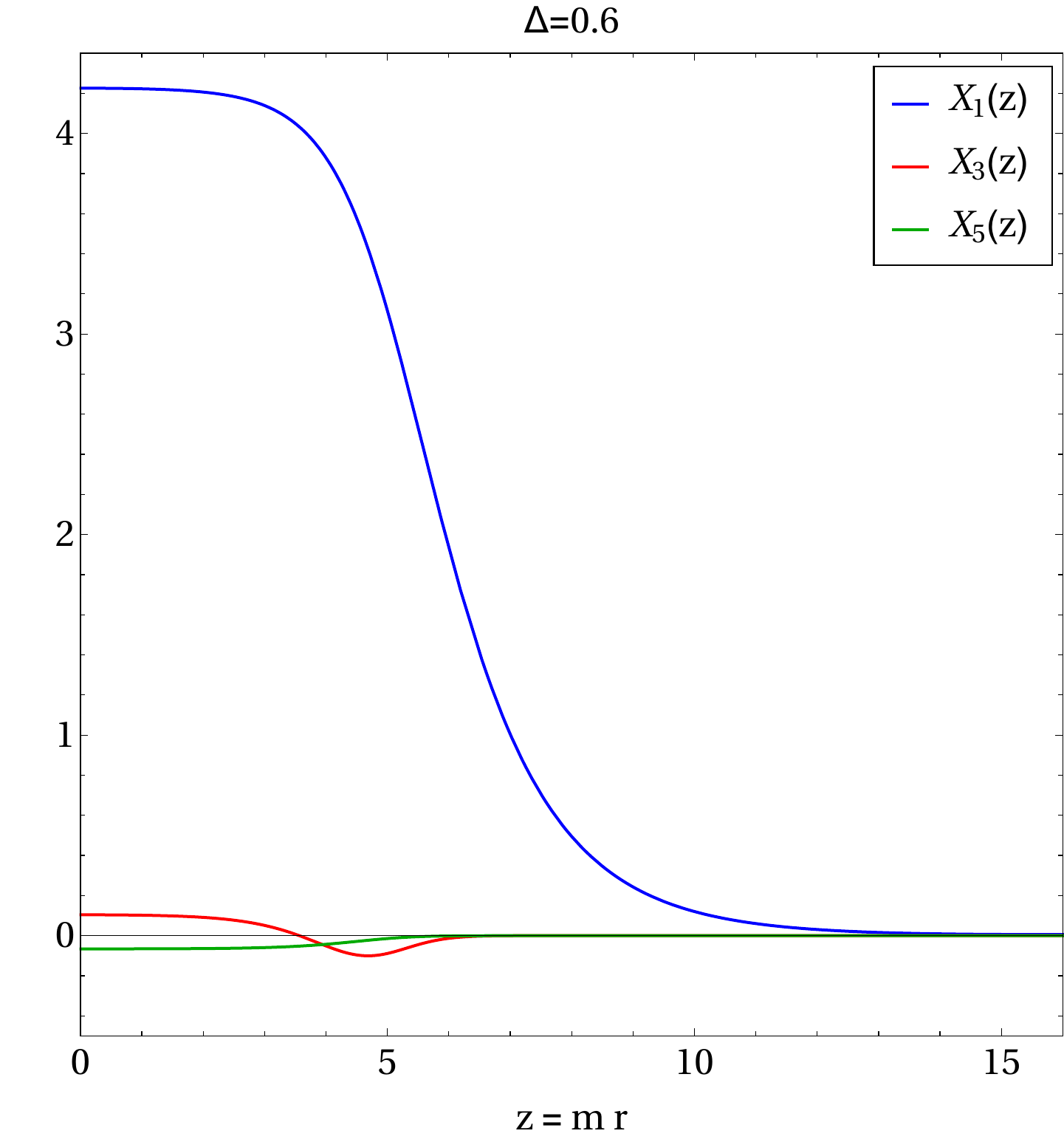}
 \caption{Solution of eqs. (\ref{Y1}), (\ref{Y3}), and (\ref{Y5}) (scaled using $X_k = \Delta^k\,Y_k$) for the parameter choice $\Delta=0.02$ (left) and $\Delta=0.6$ (right). These two configurations have nearly the same mass ${M \approx 4000 f^2/m}$ but they lie on the transition branch (left) and dense branch (right).}
 \label{DenseSolutions}
\end{figure}

Once eq. (\ref{Y1}) is solved for the leading wavefunction $Y_1$, then $Y_3$ and $Y_5$ can be solved directly. We plot the result for $X_1$, $X_3$, and $X_5$ for two parameter choices ($\Delta=.02$ and $\Delta=0.6$) in Figure \ref{DenseSolutions}; these two example points correspond to the circles on Figure \ref{Mass_Rad}. For illustration, the solutions in the left and right panels were chosen to have close to the same mass, $M\approx4000 f^2/m$, but they lie respectively on the transition and dense branches of the mass/radius curve (see Figure \ref{Mass_Rad}). We observe that the contribution of $X_3$ is much more significant at larger $\Delta$ (the right panel), and it is nonzero in an $\mathcal{O}(1)$ fraction of the volume of the star. The wavefunction is more compact and more concave for the solution in the dense branch.

We wish to calculate the physical parameters from the numerical solutions. The equation of motion for $Y_1$, eq. (\ref{Y1}), can be derived from an effective Lagrangian
\begin{equation} \label{LagrDense}
 L = \frac{1}{2}\int d^3x\left[Y_1'{}^2 + Y_1{}^2 - \frac{1}{16}Y_1{}^4
	      + \frac{\Delta^2}{512}Y_1{}^6\right].
\end{equation}
This allows a simple calculation of the effective mass at the same order in $\Delta$,
\begin{equation} \label{MassTot}
  M = \frac{\pi\,f^2}{\Delta\,m}\,\int dx\,x^2\left[(2-\Delta^2)Y_1{}^2 
	      +\Delta^2\,Y_1'^2
	      -\frac{\Delta^2\,Y_1{}^4}{16} 
	      + \frac{\Delta^4\,Y_1{}^6}{512}\right].
\end{equation}
This is, of course, the same as the expectation value 
of eq. (\ref{MassT00}) expanded to the same order. As we have pointed out, $X_5$ does not contribute to the mass of the axion star at this order, and the contribution of $X_3$ is to modify the coefficient on the $X_1{}^6$ term.

Solving eqs. (\ref{Y1}), (\ref{Y3}), and (\ref{Y5}), we plot the resulting masses and radii as the dashed, red line in Figure \ref{Mass_Rad}, which spans the range $\Delta\in\{10^{-7},0.69\}$. At very small $\Delta$ (large $R_{99}$), we have also included the results of the previous section (equivalent to \cite{ESVW}) for the dilute branch and the corresponding maximum mass at weak binding (blue, green, and yellow curves for $f=10^9,10^{11},10^{13}$ GeV respectively); on this branch, at the very top of the graph, gravity is an important factor. On the transition branch, where the methods overlap, we find terrific agreement.

We have included solutions in Figure \ref{Mass_Rad} up to $\Delta=0.69$, although at the largest values even the $\mathcal{O}(\Delta^5)$ analysis becomes questionable. The method described here can be extended further if larger values of $\Delta$ are of interest in the future. It has been suggested \cite{Braaten,ELSW,ChavanisPT} that the masses of dense axion stars extend to very large (possibly arbitrarily large) values, increasing as $M\propto R^3$; we have indicated this suggestion by the dashed blue line at the bottom of Figure \ref{Mass_Rad}. However these analyses do not include relativistic corrections \cite{WilczekASt}, and so this conclusion is not perfectly robust. The question of whether an endpoint exists for the mass of dense axion stars remains an interesting one.

It should of course be noted also that, if dense solutions truly exist at very large masses with a radius $R\propto M^{1/3}$, then at some point gravity must become important again. Consistency with General Relativity seems to require an eventual approach of axion star solutions to the Schwarzschild radius if $M$ grows much faster than $R$. Since no current method consistently describes such a region of parameter space, we do not comment further on this scenario here.

\subsection{Stability}

In this section, we will analyze the structural stability of the solutions we have considered. To do so, we will use a generalized scaling analysis, akin to the derivation of Derrick's Theorem \cite{Derrick}. 

To see how this works, we begin with the simple case of dilute axion stars (as in Section \ref{ASTtoX1}). The equation of motion, including the gravitational interaction, is eq. (\ref{diluteEOM}). After multiplying the equation by $Y_1(x)$ and integrating over volume, we can rewrite it as $K + \mathcal{N} + G + V_4 = 0$, where
\begin{align} \label{KNGV4}
 K &= \int d^3x\, Y_1'(x)^2, &G& = \kappa \int d^3x\, b(x) Y_1(x)^2  \nn \\
 \mathcal{N} &= \int d^3x\, Y_1(x)^2,  &V_4& = -\frac{1}{8}\int d^3x\, Y_1(x)^4.
\end{align}
This equation can be derived from a Lagrangian of the form
\begin{equation}
 L = \mathcal{N} + K + \frac{1}{2}\left(G + V_4\right).
\end{equation}

Now, a rescaling of the coordinate by $x\to \lambda x$ will leave the mass\footnote{Note that this expression is the $\Delta\to0$ limit of eq. (\ref{MassTot}).}
\begin{equation}
 M_{\Delta\to 0} = \frac{f^2}{2\,m\,\Delta}\int d^3x\, Y_1(x)^2
\end{equation}
invariant as long as the wavefunction is also scaled by $Y_1\to \lambda^{-3/2}\,Y_1$. Then the Lagrangian is rescaled as
\begin{equation}
 L_\lambda = \mathcal{N} + \frac{K}{\lambda^2} 
	+ \frac{1}{2}\left(\frac{G}{\lambda} + \frac{V_4}{\lambda^3}\right)
    \equiv \mathcal{N} + E_\lambda.
\end{equation}
A solution of the equation of motion will satisfy
\begin{equation} \label{dLdlam}
 -\partial_\lambda\,E_\lambda \big|_{\lambda=1} = 2K 
	  + \frac{1}{2}G + \frac{3}{2}V_4 = 0.
\end{equation}
It is the second variation, 
\begin{equation} \label{d2Ldlam2}
  \delta^2 E_\lambda 
	\equiv \partial_\lambda{}^2 E_\lambda\big|_{\lambda=1} 
	= 6K + G + 6V_4,
\end{equation}
that signifies the structural stability of a solution; $\delta^2 E_\lambda >0$ is a necessary condition for stability, and $\delta^2 E_\lambda <0$ is a sufficient condition for instability.

\begin{figure}[t]
 \centering
 \includegraphics[scale=.53]{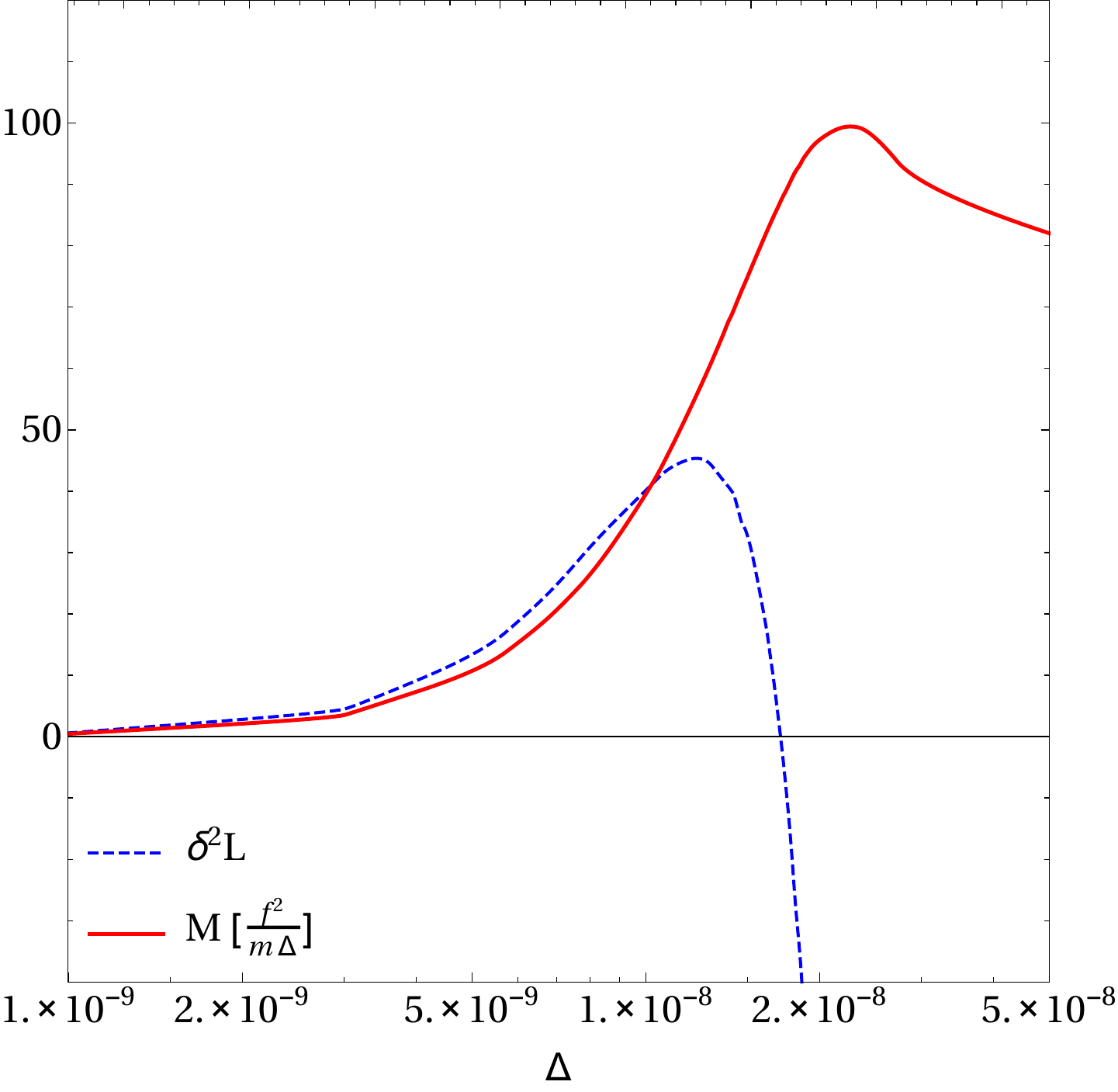} ~~~
 \includegraphics[scale=.49]{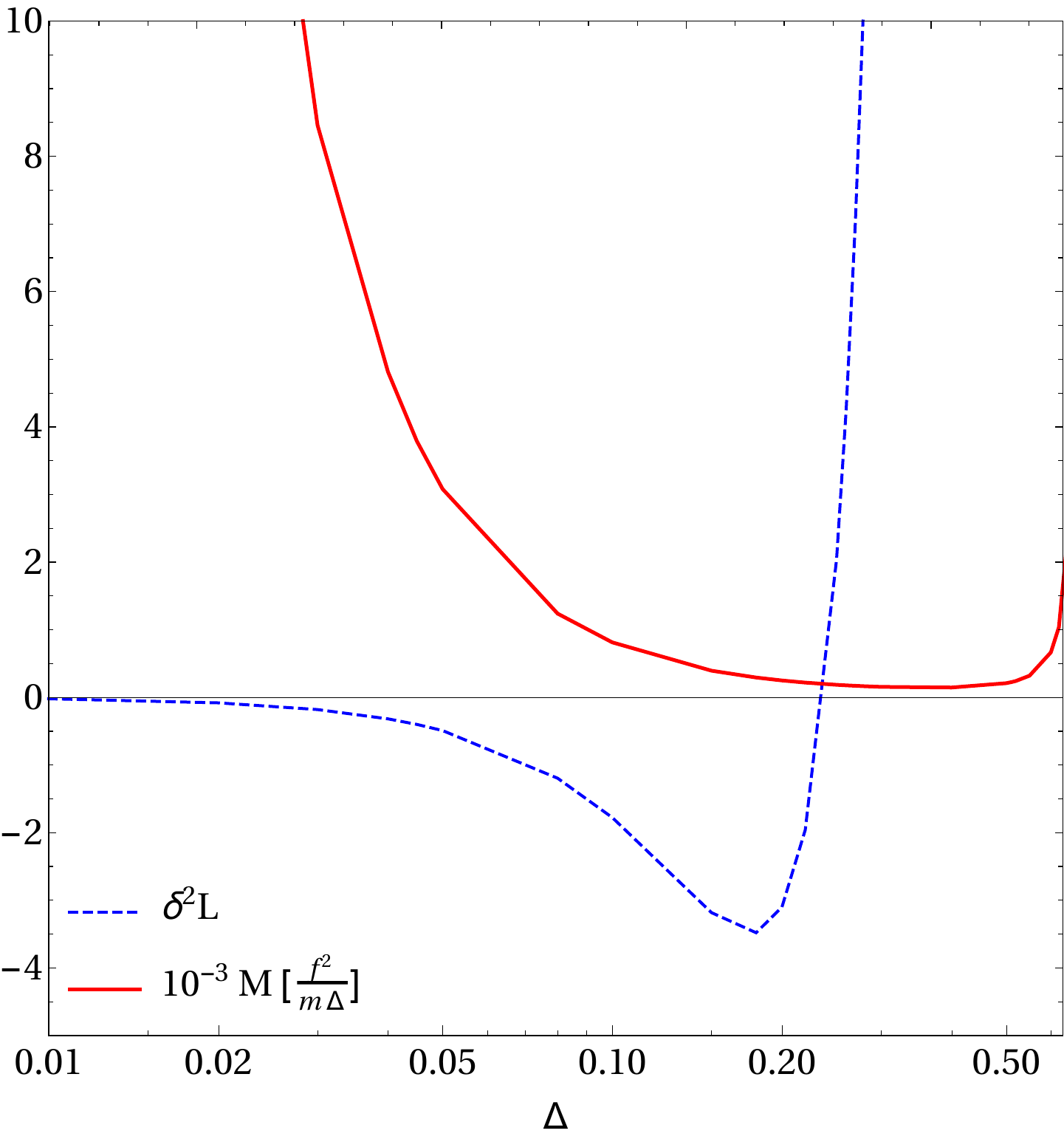}
 \caption{The stability of axion stars as determined by a perturbation of the Lagrangian. The dashed, blue curves indicate the second variation of the Lagrangian, which must be $>0$ for a solution to be stable, as a function of $\Delta$; the solid, red curves are the corresponding masses. The left panel depicts the crossover from the dilute (stable) to transition (unstable) branches at $\Delta_c \approx f/M_P$, for the parameter choice $f=10^{11}$ GeV; the right panel depicts the crossover from transition (unstable) to dense (stable) branches at $\Delta \approx 0.27$ (independent of $f$).}
 \label{Stability}
\end{figure}

We show the result of this analysis in the left panel of Figure \ref{Stability}, for the parameter choice $f=10^{11}$ GeV. It is easy to see that the crossover from stable to unstable occurs at the maximum mass, with smaller values of $\Delta$ being stable. This occurs around $\Delta \approx 2\times10^{-8}$, though the numerical value is dependent on the choice of $f$; more generally, the crossover occurs near $\Delta_c \approx f/M_P$. Of course we also verified that the solutions of eq. (\ref{diluteEOM}) satisfy $\delta L = 0$ to very high precision.

It is quite instructive to investigate the stability of two special cases. If we set $G = 0$, then we consider self-interacting bosons without gravity. Then eq. (\ref{dLdlam}) implies $V_4 = -4 K / 3$. Substituting this into the second variation in eq. (\ref{d2Ldlam2}), we obtain $\delta^2 E_\lambda = -2 K < 0$. Consequently, no solution of the equation of motion is stable. This is just Derrick's theorem directly applied to boson stars \cite{Derrick}. This situation also applies to the majority of states along the transition branch of axion stars; in that region gravity has decoupled, but higher-order self-interactions have not yet become relevant. In this region, we find that the solutions are unstable, in agreement with other works \cite{ChavanisMR,ChavanisMR2}.

Another special case is bosons with gravitational interactions only. This is the case studied by Ruffini and Bonazzola \cite{RB}. Then eq. (\ref{dLdlam}) gives $G = -4 K$. Substituting this into eq. (\ref{d2Ldlam2}) we obtain $\delta^2 E_\lambda = 2 K > 0$. In other words, every solution of eq. (\ref{diluteEOM}) is stable if we neglect the $X_1{}^3$ term completely.\footnote{This conclusion breaks down at large masses where relativistic corrections become important; as the original works on boson stars indicate \cite{Kaup,RB}, even noninteracting theories have a maximum mass for boson stars, which is $M_{max}^{NI}=0.63\,M_P{}^2/m$. Of course, with GRB we recover this result as well.}

We may apply a similar analysis to our GRB solutions to the system (\ref{Y1}). We begin by rewriting eq. (\ref{Y1}) as $K + \mathcal{N} + V_4 + V_6 = 0$, using eq. (\ref{KNGV4}) and defining
\begin{equation}
 V_6 = \frac{3}{512}\int d^3x\,Y_1(x)^6.
\end{equation}
Then the total mass in eq. (\ref{MassTot}) is
\begin{equation}
 M = \frac{f^2}{4\,m\,\Delta}\left[2\,\mathcal{N} 
	+ \Delta^2\left(K - \mathcal{N} + \frac{1}{2}V_4\right)
	+ \frac{\Delta^4}{3}V_6\right].
\end{equation}
Under a rescaling $x\to\lambda\,x$, it is no longer so straighforward to rescale $Y_1\to \mu\,Y_1$ in a way which leaves $M$ invariant. 

We can however derive scaling relations order by order in $\Delta$. We will define $\lambda$ as a power series in $\Delta$,
\begin{equation}
\lambda=\lambda_0+\Delta^2\,\lambda_2+\Delta^4\,\lambda_4,
\end{equation}
where $\lambda_i$ are dependent on $\mu$. Then after applying $x\to\lambda\,x$ and $Y_1\to \mu\,Y_1$, we can write
\begin{align}\label{mass3}
 \frac{4\,m\,\Delta}{f^2} M &= \lambda_0{}^3\,\mu^2\,2\,\mathcal{N} \nn \\
	&+ \Delta^2\left[6\,\lambda_0^2\,\lambda_2\,\mathcal{N}\,\mu^2
		+ \lambda_0\,\mu^2\,K-\lambda_0{}^3\,\mu^2\,\mathcal{N}
		+\frac{\lambda_0{}^3\,\mu^4}{2}\,V_4 \right]	\nn	\\
  &+ \Delta^4\left[(2\,\lambda_2{}^2\,\lambda_0
	  - \lambda_0{}^2\lambda_2+2\,\lambda_0{}^2\,\lambda_4\,)\,3\,\mathcal{N}\,\mu^2
	  + \lambda_2\,\mu^2\,K 
	  + \frac{3\,\lambda_0{}^2\,\lambda_2\,\mu^4}{2}\,V_4
	  + \frac{\lambda_0{}^3\,\mu^6}{3}\,V_6 \right].
\end{align}
Then we can solve order by order for the components of $\lambda$, using the following equations:
\begin{align}
  \lambda_0{}^3\,\mu^2 &= 1,	\nn	\\
  \lambda_0\,\mu^2\left[K 
	+ \left(6\,\lambda_0\,\lambda_2 - \lambda_0{}^2\right)\mathcal{N}
	+ \frac{\lambda_0{}^2\,\mu^2}{2}V_4\right] 
		  &= K - \mathcal{N} + \frac{1}{2}V_4	\nn	\\
  \mu^2\left[\lambda_2\,K
	+ 3\left(2\,\lambda_0\,\lambda_2{}^2 + 2\,\lambda_0{}^2\,\lambda_4
			  - \lambda_0{}^2\,\lambda_2\right)\mathcal{N}
	+ \frac{3\,\lambda_0{}^2\,\lambda_2\,\mu^2}{2}V_4
	+ \frac{\lambda_0{}^3\,\mu^4}{3}V_6\right] &= \frac{1}{3}V_6.
\end{align}
These equations are solved as
\begin{align}\label{lambdas}
 \lambda_0 &= \mu^{-2\,/\,3}, 	\nn	\\
 \lambda_2 &= \frac{\mu^{-2\,/\,3}}{12\,\mathcal{N}}\,
	\left[\left(1-\mu^{4\,/\,3}\right)\,2K + \left(1-\mu^2\right)V_4\right],\nn\\
 \lambda_4 &= \frac{\mu^{-2\,/\,3}}{144\,\mathcal{N}^2}
      \Bigg[K\left(12\,\mathcal{N} - 4\,V_4 
		    + 2\,\mu^{4\,/\,3}\left(V_4-6\,\mathcal{N}\right)
		    - 2\,\mu^2\,V_4 + 4\,V_4\,\mu^{10\,/\,3}\right)
		    + 4\,K^2\left(\mu^{4\,/\,3}-1\right) \nn \\
	    &~~~ 
		      + \mathcal{N}\left(6\,V_4 + 8\,V_6 - 6\,\mu^2\,V_4 
		      - 8\,\mu^4\,V_6\right)
	    + V_4^2\left(2\,\mu^4 - \mu^2 - 1\right)\Bigg].
\end{align}
Under a rescaling of this kind, $M$ remains constant under rescaling at $\mathcal{O}(\Delta^4)$. 

As before, after rescaling the Lagrangian of eq. (\ref{LagrDense}), we verify that ${\delta L \equiv \partial_\mu\,L\big|_{\mu=1} = 0}$ for our solutions. We plot the second variation $\delta^2 L$ in Figure \ref{Stability} and find that above $\Delta \approx 0.27$, the solutions are again structurally stable. This corresponds to the crossover from the transition branch to the dense branch of axion star solutions, and is very near to the position of the local minimum of the mass, which is roughly $M_{min}\approx 463\,f^2/m$.

Previous works using the single-harmonic equation of motion have concluded that states on the dilute and dense branches are structurally stable, whereas states on the transition branch are unstable \cite{ChavanisMR,ChavanisMR2,ELSW,WilczekASt}. We conclude from the perturbation analysis of our solutions at $\mathcal{O}(\Delta^5)$ that this result is robust; this is important because, on the dense branch, the stability might have been affected by the higher order corrections we calculated here. We have only considered perturbations on $Y_1$, whereas a more general analysis would require independent perturbations also on $Y_3$ and $Y_5$; there could be directions in this space which are unstable. We leave such a general analysis to a future work.

\section{Conclusion} \label{Conclusion}

Previous work on boson stars has been dependent on the (often tacit) assumption that the binding energy of the constituent bosons is small. Such stars are well described by a nonrelativistic analysis where the field has a single harmonic dependence on the particle eigenenergy $E_0$. However, as stars become denser, their binding energy grows and corrections to this simple picture become important. A related problem is that for self-interacting theories, the operator Klein-Gordon equation is not exactly satisfied by a single wavefunction with harmonic time dependence.

Building on the seminal work of Ruffini and Bonazzola \cite{RB}, we have presented a general analysis method which solves these problems. The Generalized Ruffini-Bonazzola (GRB) operator in eq. (\ref{GRB}), is an iterative solution for bound states of the operator Klein-Gordon equation with arbitrary self-interaction; in principle, this formalism includes all relevant corrections in special relativity. The solution can be viewed as a tower of expectation values of the Klein-Gordon equation in eq. (\ref{Mk}), which are solved for a series of wavefunction components $R_k$  with harmonic time dependence of energy $k\,E_0$.

In principle, the full solution requires one to solve $N\ggg1$ equations for $N\ggg1$ wavefunctions. In practice, this procedure is tractable because the tower of equations (\ref{Mk}) can be viewed as a power series in a small parameter $\Delta = \sqrt{1 - E_0{}^2/m^2}$. In the very weak binding limit $\Delta\ll 1$, the leading order GRB field operator of eq. (\ref{GRB}) reduces to the single harmonic case of eq. (\ref{RB}), and as such the expansion justifies the truncation at this order. Corrections come at higher order in $R_k\propto\Delta^k$ and can be calculated iteratively to whatever order is appropriate. In Section \ref{SolBS}, we illustrated the procedure using the computation of $\mathcal{O}(\Delta^5)$ calculation for a boson star with quartic self-interactions.

We also applied the GRB framework to the case of the axion potential, eq. (\ref{Vaxion}). The leading attractive self-interaction enters at $\mathcal{O}(\Phi^4)$, though the full potential contains in an infinite series of interaction terms with alternating signs.\footnote{In the case of the chiral axion potential \cite{Vecchia,Cortona}, the signs do not exactly alternate, but we ignore this complication here.} Solutions to the equations of motion for an axion star enter on three distinct branches: ``dilute'', where gravity is important; ``transition'', where gravity decouples and the leading self-interaction is dominant; and ``dense'', where higher order self-interactions become of comparable order. These three regions can be viewed as different limits of the parameter $\Delta$:
\begin{align}
 &\Delta \lesssim \frac{f}{M_P} \qquad &(\text{Dilute Branch}) \nn \\
 \frac{f}{M_P} \lesssim &~\Delta \ll 1 \qquad &(\text{Transition Branch}) \nn \\
 &\Delta = \mathcal{O}(1) \qquad &(\text{Dense Branch}). \nn
\end{align}
Of course, $\Delta<1$ always obtains for axions which are bound to the axion star.

At the order we have computed here, the differential equation has a single free parameter, which can be chosen as $M$, $\Delta$, central density $Y_1(0)$, etc. We choose to set $\Delta$ by hand as a numerical convenience, because it is used as an expansion parameter and because all physical quantities are one-to-one functions of $\Delta$. Once $\Delta$ is fixed, there is a single $Y_1(0)$ which gives a solution that goes asymptotically to $0$ as $x\to\infty$, and the mass is also fixed by this solution. To the extent that $N$ is a reasonable physical quantity (at least at weak binding), it is easy to calculate from the solution as well. These physical quantities are all mutually dependent.

Expanding the axion potential to the leading \emph{repulsive} order, $\mathcal{O}(\Phi^6)$, we analyzed the crossover between the transition and dense branches of axion stars. In the context of GRB this implies evaluation up to $\mathcal{O}(\Delta^5)$ and solution for three wavefunction components $R_1$, $R_3$, and $R_5$. We found with high precision the position of the crossover at $\Delta \approx 0.27$, which corresponds to a local minimum of the mass at $M_{min} \approx 463\,f^2/m$. A full characterization of the dense branch to very large masses would require an even higher order expansion, which is beyond the scope of this work.

Finally, we have examined the stability of axion star solutions using a perturbation analysis of the leading order wavefunction $R_1$. We confirm that structural stability is obtained on the dilute and dense branches of solutions, whereas the transition states are unstable. A more general analysis would require independent perturbation of the higher-order component wavefunctions $R_{k>1}$, a task which we postpone until future work.

We have indicated throughout that the bound state solutions to the Klein-Gordon equation should be supplemented by a collection of scattering states as well. Such scattering states give rise to decay processes by which bound bosons are converted to relativistic bosons through self interactions \cite{ESW,EMSW}. A description of these scattering states can be obtained consistently within the GRB framework, and as with the bound state solutions, the decay matrix elements for boson stars can be calculated iteratively as a power series in $\Delta$. We will describe this further in a companion publication, to appear soon.

\section*{Note Added}

While this work was being finalized, a partially-overlapping work appeared \cite{Guth2}. Their approach to relativistic corrections is based on the Gross-Pitaevskii equation, whereas ours is based on an extension of Ruffini-Bonazzola; we thus believe these methods are likely to be complementary.

\section*{Acknowledgements}
 
We thank M. Amin, E. Braaten, H. Kim, M. Ma, A. Mohapatra, M. Takimoto, G. Tavares, H. Zhang for fruitful discussions.  The work of J.E. was supported by the Zuckerman STEM Leadership Program. J.E. also thanks the Fermilab Theory Department, where some of this work was completed, for their hospitality.

\end{document}